\documentclass[journal]{vgtc}                     

\onlineid{0}

\vgtccategory{Research}

\title{Understanding Organizational Strategies Across \\ Multimodal Artifacts in Immersive Computational Notebooks}

\author{%
  \authororcid{Sungwon In}{0000-0002-5316-2922},
  \authororcid{Minju Baeck}{0000-0001-7179-2103},
  \authororcid{Yalong Yang}{0000-0001-9414-9911},
  \authororcid{Sang Ho Yoon}{0000-0002-3780-5350},
  \authororcid{Woontack Woo}{0000-0002-5501-4421}, and
  \authororcid{Mallesham Dasari}{0000-0002-8855-036X}
}

\authorfooter{
  \item
  	Sungwon In is with Kyung Hee University.
  	E-mail: sungwon.i@khu.ac.kr
  \item
  	Minju Baeck is with KAIST.
  	E-mail: minjubaeck@kaist.ac.kr.
  \item
  	Yalong Yang is with Georgia Tech.
  	E-mail: yalong.yangi@gatech.edu.
  \item
  	Sang Ho Yoon is with KAIST.
  	E-mail: sangho@kaist.ac.kr.
  \item
  	Woontack Woo is with KAIST.
  	E-mail: wwoo@kaist.ac.kr.
  \item
  	Mallesham Dasari is with Northeastern University.
  	E-mail: m.dasari@northeastern.edu.
}

\abstract{%
Immersive Computational Notebooks (ICoN) extend traditional notebook environments into immersive spaces, enabling analysts to interact with multimodal artifacts, including code, narratives, data tables, and visualizations. 
By integrating multimodal artifacts into a single immersive workspace, ICoN enables analysts to transition between analytical tasks seamlessly.
Meanwhile, understanding organizational strategies is critical for designing effective interactions to further support analysts.
However, prior research on immersive computational notebooks has primarily examined organizational strategies centered on single-modality artifacts. 
Systematic investigations of how analysts spatially organize the complex relationships among multimodal artifacts in a single immersive workspace remain underexplored.
To address this gap, we conducted a user study to examine organizational strategies for multimodal artifacts in immersive computational notebooks. 
Our findings show that participants predominantly adopted depth-based layouts, and their spatial organization was largely structured around cell-based artifacts.
}


\teaser{
    \centering
    \includegraphics[width=1\textwidth]{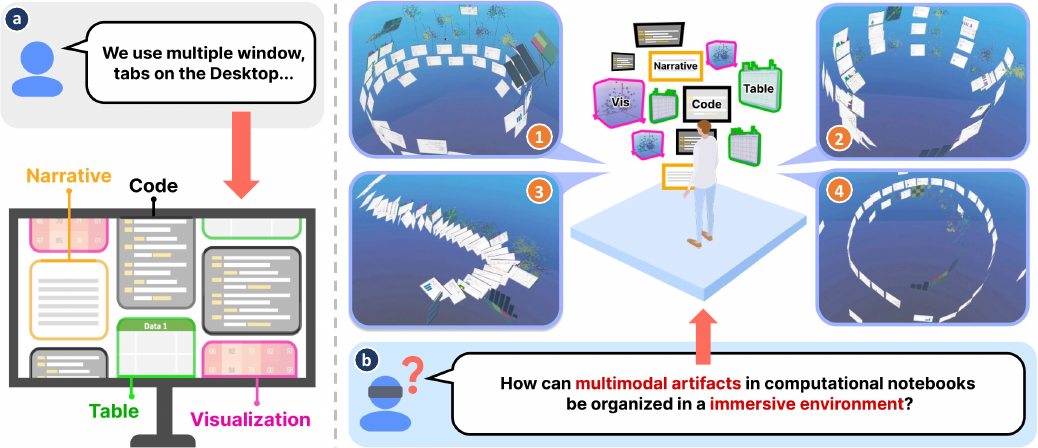}
    \caption{(a) In a 2D desktop, users organize multimodal artifacts across multiple windows. (b) Users then wonder how the multimodal artifacts in computational notebooks can be organized in an immersive environment. (1)-(4) illustrate different organizational strategies: (1) artifacts organized in multiple horizontal layers, (2) in vertical layers, and (3) and (4) organized along the depth axis.}
    \vspace{2mm}
    \label{fig:teaser}
}

\graphicspath{{figs/}{figures/}{pictures/}{images/}{./}} 

\usepackage{booktabs}                  
\usepackage{lipsum}                    
\usepackage{mwe}                       
\usepackage{ccicons}                   

\usepackage{mathptmx}                  

\usepackage{tabu}                      
\usepackage{booktabs}                  
\usepackage{lipsum}                    
\usepackage{mwe}                       

\usepackage[english]{babel}
\addto\extrasenglish{%
}

\AtBeginDocument{%
  }

\usepackage{subfigure}
\usepackage{makecell}

\usepackage{enumitem}
\setitemize{noitemsep,topsep=0pt,parsep=0pt,partopsep=0pt}

\usepackage{hyphenat}

\usepackage{forloop}

\usepackage[defaultcolor=black]{changes}

\newsavebox\MyBreakChar%
\sbox\MyBreakChar{}
\newsavebox\MySpaceBreakChar%
\sbox\MySpaceBreakChar{\hyp}
\makeatletter%
\newcommand*{\BreakableChar}[1][\MyBreakChar]{%
  \leavevmode%
  \discretionary{\usebox#1}{}{}%
}%
\makeatother

\newcounter{index}%
\newcommand{\AddBreakableChars}[1]{%
  \StrLen{#1 }[\stringLength]%
  \forloop[1]{index}{1}{\value{index}<\stringLength}{%
    \StrChar{#1}{\value{index}}[\currentLetter]%
    \IfStrEqCase{\currentLetter}{%
        {*}{\currentLetter\BreakableChar[\MyBreakChar]}%
        {/}{\currentLetter\BreakableChar[\MyBreakChar]}%
        {+}{\currentLetter\BreakableChar[\MyBreakChar]}%
        {\&}{\currentLetter\BreakableChar[\MyBreakChar]}%
    }[\currentLetter]%
  }%
}%

\def\fixedB{\raisebox{-0.1\height}{\includegraphics[height=0.77em]{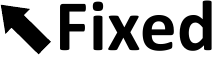}}}
\def\hybridB{\raisebox{-0.22\height}{\includegraphics[height=0.88em]{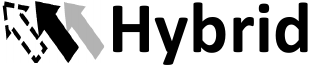}}}

\def\fixedT{\textsc{Fixed}}
\def\hybridT{\textsc{Hybrid}}

\begin{document}



\maketitle

\section{Introduction}
\label{sec:intro}


The rapid development of immersive technologies has introduced new paradigms across various domains~\cite{baeck2025visuo, reiske2023multi, zhang2025forcepinch}. 
Notably, immersive data science is an emerging field that leverages novel interaction and large display to enhance data analytics~\cite{cordeil2019iatk, in2023table}.
Specifically, analysts can organize artifacts in a large 3D space, which enables them to externalize analytical intent more intuitively~\cite{davidson2022exploring}.

Meanwhile, organizational strategies in immersive space vary depending on the analysis being performed and the artifact being manipulated. 
\added{For instance, when analyzing data to uncover insights, data tables or visualizations are often positioned close together to support side-by-side comparison~\cite{reiske2023multi, tong2025exploring}.}
In contrast, code artifacts in immersive space are often organized according to execution logic, such as branching structure, when analysts test multiple hypotheses with different analytical paths~\cite{in2024evaluating}. 
Beyond immersive workspaces that support only a single modality (e.g., either data or code), Immersive Computational Notebooks (ICoN) extend the computational notebook paradigm into immersive space, enabling analysts to work with multimodal artifacts through a unique proxy called a cell~\cite{in2025investigating}.
\added{Within an immersive computational notebook, analysts execute code cells to perform computations, and pull the resulting 2D data table or visualization as a 3D projected data artifact to perform more intuitive and visual-based exploration.}

\added{Most importantly, the existence of multimodal artifacts within a single immersive workspace begins to create unique organizational possibilities.}
For instance, 3D projected data artifacts can be placed in front of code, forming a depth-oriented organization.
\added{However, such an organization may result from the designed interaction mechanism (e.g., pull-out) rather than from users' perceived relationships between cells and data artifacts.
Although providing an organization that clearly conveys analytical intent can help preserve complex workflows and reduce the effort required to manage growing analysis, prior work has largely focused on single-modality organizational strategies~\cite{in2025exploring, luo_documents_2025}.
Furthermore, while computational notebooks often incorporate narratives to communicate analytical intent~\cite{pimentel2019large}, how narratives shape organizational strategies alongside code and data artifacts in immersive space has not been explored.}
\textbf{Therefore, our first goal is to understand organizational strategies in multimodal immersive computational notebooks, focusing on code, narrative, data table, and visualization artifacts.}

Building on our first goal, organizational strategies are often shaped by how analysts establish and manage flows~\cite{ren2019understanding}.
Flows clarify relationships and dependencies in analysis and are commonly represented with explicit line-based encodings~\cite{telea1999vission, yu2016visflow}.
\added{Similarly, in immersive computational notebooks, flows are shown as explicit directional lines.
However, analysts may rely on spatial proxies, such as placing related artifacts close together without explicit links, as numerous explicit lines can introduce visual clutter and occlusion~\cite{in2025exploring, yang2016many}.
Meanwhile, long-standing efforts have explored implicit flow representations (e.g., transparent lines) to balance expressiveness and visual simplicity~\cite{zhou2013edge}.
\textbf{Therefore, our second goal is to investigate how flow representation influences organizational strategies in multimodal immersive computational notebooks.}}

\added{To address our research goals, we conducted a controlled user study with two conditions: Fixed, in which only explicit flow indicators were available, and Hybrid, in which participants could use explicit, implicit flows, and spatial proxies, enabling more flexible representations of flow.
In both conditions, participants organized four types of artifacts: codes, narratives, data tables, and visualizations.
Participants were asked to organize in a way that made sense to them. 
Our results reveal that participants largely favored depth-based layouts.} 
In addition, we observed unique flow between multimodal artifacts. 
Non-linear execution orders were not dominant among cell-based artifacts, and flows predominantly originated from cell-based artifacts toward data-based artifacts.
Furthermore, in Hybrid, we observed that most participants continued to use explicit lines between cells; they preferred only a spatial proxy to represent flows between cells and data artifacts.
Therefore, we suggest supporting depth as a potential organizational dimension and context-aware flow representations to facilitate interaction with multimodal immersive computational notebooks.

\textbf{In Summary, the contributions of this paper are twofold:} \textit{first,} a systematic exploration of organizational strategies for multimodal immersive computational notebooks; \textit{and second,} design implications to further support the use of multimodal immersive computational notebooks.

\section{Related Work}
\label{sec:relwork}

\textbf{Artifacts in Computational Notebooks.}
Drawing from Knuth’s literate programming paradigm~\cite{knuth1984literate}, computational notebooks construct an analytical pipeline using a unique proxy called the cell, which integrates code, outputs, and narratives into a unified document.
In desktop-based notebooks (e.g., Jupyter~\cite{JupyterNotebook}, Observable~\cite{Observable}), analysts work with four primary types of cell-based artifacts: 1. code-only cells; 2. code cells that contain data table outputs, where loaded data are previewed in tabular form; 3. code cells that contain data visualizations; and 4. narrative cells, used to record contextual information.
However, working with multiple artifacts in a 2D desktop environment introduces considerable challenges due to the limited size of the workspace~\cite{chattopadhyay2020s}. 
Text-based artifacts must be navigated linearly through extensive scrolling~\cite{head2019managing, kery2017variolite, rule2018exploration}, making large-scale analyses difficult to manage, while data artifacts are limited in their support for dynamic exploration~\cite{yang2020tilt, yang2022pattern}.
To address these limitations, In et al. introduced the Immersive Computational Notebook (ICoN), which extends the notebook paradigm into immersive space~\cite{in2025investigating, in2025towards}.
In ICoN, code and narrative cells can be arranged non-linearly in 3D space, while data output can be extracted from cells and instantiated as manipulable 3D objects for more intuitive exploration.
Although ICoN enables richer interaction with multimodal artifacts in immersive environments, it has not systematically examined how analysts organize the complex relationships that emerge when multiple artifact types coexist within a single workspace.

\textbf{Spatial Organization in Immersive Environments.}
Organizations are a representation of how users understand a given task and how they externalize their reasoning processes~\cite{rule2018exploration}.
In immersive environments, users have been shown to adopt different organizational strategies depending on the types of artifacts. 
For instance, when performing analytical tasks with text-based artifacts (e.g., code~\cite{in2025exploring} or PDF documents~\cite{liu2020design, satriadi2020maps}), artifacts are often organized according to the employed flow, reflecting the sequence of execution or analytical reasoning.
In contrast, when working with data-based artifacts (e.g., data tables~\cite{in2023table} or visualizations~\cite{reiske2023multi, cordeil2019iatk}), users have tended to place artifacts in close proximity, such as side-by-side~\cite{harden2022exploring}, to facilitate inspection and comparison across datasets, rather than organizing them to reflect the sequence.
Notably, regardless of the artifact types, users have tended to form circular layouts surrounding their bodies, creating workspaces that facilitate engagement~\cite{in2025exploring}.
In addition to artifact types, organizational strategies also vary depending on environmental context. 
For instance, in augmented reality (AR) settings, users’ organization is often influenced by surrounding physical objects, such as avoiding the placement of virtual artifacts near fragile real-world objects (e.g., monitors or glasses)~\cite{luo_documents_2025}.
However, most prior work has focused on organizational strategies involving single-modality artifacts within a given workspace (e.g., only code, PDF documents, or data), which has not addressed how complex relationships among multimodal artifacts can be externalized through spatial organization.
Therefore, we are interested in exploring how analysts organize multimodal artifacts when multiple artifact types are present within a single immersive workspace.

\textbf{Flow Diagram in Data Science.}
In data science, flow diagrams are widely used to explicitly represent analytical workflows through proxies such as nodes and edges~\cite{IBMSPSSModeler, KNIME, wolstencroft2013taverna}. 
Nodes denote tasks to be executed or operations to be performed, while edges indicate how work proceeds from one node to another~\cite{bergmann2011retrieval}. 
For instance, in a desktop environment, Domino enables users to extract, compare, and refine subsets across multiple tabular datasets, making intermediate data artifacts explicit and supporting exploration without losing provenance~\cite{gratzl2014domino}. 
Similarly, VisFlow visualizes how data streams move between operations and visual outputs, allowing users to trace how each transformation affects downstream artifacts~\cite{yu2016visflow}.
Similar logic appears in multimodal immersive computational notebook, where nodes are represented as cells or data-based artifacts, and flow indicators function as edges to express execution order between cells or data flow between cell–data and data–data artifacts~\cite{in2025investigating}.
However, as the number of artifacts and flow indicators increases, workspaces can become visually dense, leading to cluttered layouts and reduced comprehensibility~\cite{in2025exploring, yoghourdjian2020scalability}.
To improve visual clarity, prior work has investigated adjustments to the visual properties of flow indicators, including transforming explicit indicators into more implicit representations by adjusting transparency, line width, and curvature~\cite{cui2008geometry, dickerson2002confluent, gansner2006improved, wong2003edgelens}.
While implicit visual indicators are effective for managing dense and numerous flows, their application has not been explored in the context of multimodal immersive computational notebooks.
Motivated by these observations, we aim to investigate how implicit representation of flow indicators influences users’ organizational strategies in multimodal immersive computational notebooks.

\section{User Study}
\label{sec:usability_study}

The primary objective of our study is to explore organizational strategies for multimodal immersive computational notebooks, including code, narratives, data tables, and visualizations.
Broadly speaking, organization in immersive computational notebooks is shaped both by how artifacts are positioned and by how flows are expressed. 
Specifically, flows denote the perceived relationships and execution dependencies among artifacts, which influence how analysts sequence tasks, cluster related artifacts, and organize the overall workspace.
In addition, organizations may be further reshaped depending on the representation of flow indicators~\cite{cui2008geometry}.
To this end, we aim to address the following three research questions:

\textbf{RQ-1. How are multimodal artifacts positioned in immersive computational notebooks?}
Analysts adopt different organizational strategies depending on the types of artifacts involved and the analytical goals being pursued. 
However, prior work has largely focused on how individual artifacts---such as code~\cite{in2025exploring}, narratives~\cite{luo_documents_2025}, data tables~\cite{in2023table}, and visualizations~\cite{cordeil2017imaxes}---are positioned and how these placements form specific layouts.
Therefore, our study examines how analysts position multiple artifact modalities within a single immersive workspace.

\textbf{RQ-2. What is the established flow in multimodal immersive computational notebooks?}
Flows in data science are often interpreted as how analysts reason about their analyses and how analytical processes proceed over time~\cite{lisle_sensemaking_2021}.
Accordingly, analysts establish various flows depending on how they perceive and interpret a given task~\cite{yu2016visflow}. 
In multimodal immersive computational notebooks, the presence of multimodal artifacts can introduce complex flows; yet it remains unclear how such flows are established.
Therefore, we aim to identify how flows are represented in multimodal immersive computational notebooks.

\textbf{RQ-3. How does the representation of flow indicators impact organizational strategies?}
Flows are often represented using explicit visual indicators such as arrows or connections~\cite{yu2016visflow, yu2019flowsense}.
However, representing complex relationships among multimodal artifacts using numerous explicit flow indicators can introduce visual clutter~\cite{cui2008geometry}, which may prompt participants to adjust their organizational strategies to reduce clutter~\cite {in2025exploring}.
Meanwhile, providing flexible representations of flow indicators may help participants mitigate visual clutter.
Therefore, we investigate how the representation of flow indicators influences organizational strategies in multimodal immersive computational notebooks.

\begin{figure}
    \centering
    \includegraphics[width=0.9\columnwidth]{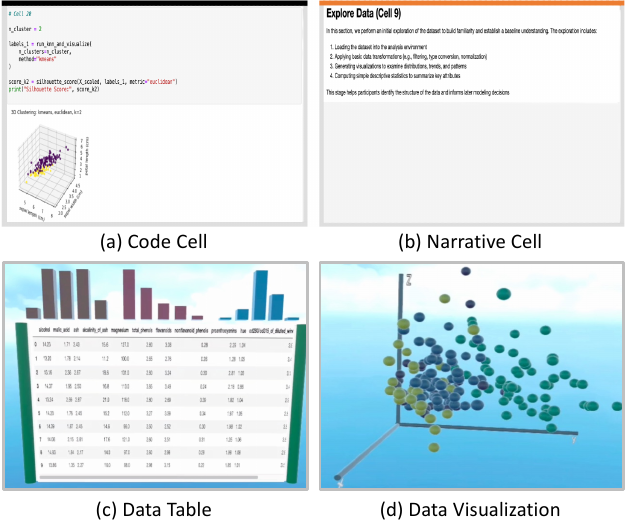}
    \caption{Four artifacts employed in our user study.}    
    \label{fig:artifact}
\end{figure}

\subsection{Study Setup}

\textbf{Multimodal artifacts.} 
Conventional desktop-based computational notebooks use cell-based artifacts that contain code, narratives, and data outputs organized along a two-dimensional axis~\cite{harden2023sage3, JupyterNotebook, Observable}.
In contrast, multimodal immersive computational notebooks extend the 2D paradigm into 3D spaces, enabling users to engage with data output (e.g., data tables and visualizations) more intuitively through manipulation of 3D projected data artifacts~\cite{in2025investigating}. 
To reflect these characteristics, our study included four artifacts:  two cell-based artifacts (code and narrative) and two data-based artifacts (data tables and visualizations). 
For the cell-based artifacts, we followed methodologies from prior computational notebook research by capturing screenshots of each code and narrative cell and placing them in movable 2D windows~\cite{harden2022exploring} (see~\autoref{fig:artifact}-(a) and (b)).
Also, code cells were intentionally made non-executable to allow participants to focus on organization rather than coding or debugging.
For the data-based artifacts, we generated 3D renderings of data tables and visualizations corresponding to the outputs produced by the associated code cells (see~\autoref{fig:artifact}-(c) and (d)).

\textbf{Number of artifacts.} 
We then considered the total number of artifacts to include in the study based on prior work and real-world notebook usage. 
Prior studies on immersive computational notebooks suggest that approximately 50 artifacts require intensive effort while keeping study length manageable~\cite{in2025towards}.
To guide the distribution of artifact types, we first considered how many cells should be included within 50.
For cell-based artifacts, we drew on findings by Pimentel et al., who reported that approximately 25\% of notebook cells produce data outputs (e.g., tables or visualizations), while fewer than 10\% contain only narrative text~\cite{pimentel2019large}.
For data-based artifacts, we followed a common pattern in immersive computational notebooks where analysts convert 2D cell outputs into 3D artifacts for further inspection when in-cell representations are insufficient~\cite{chattopadhyay2020s}; therefore considered a one-to-one correspondence between output-producing cells and 3D data-based artifacts.
Following this, we initially created 40 cell-based artifacts (26 code-only, 10 code-with-output, and 4 narrative) and 10 data-based artifacts.
However, we identified that this configuration resulted in a cell-dominant organization that failed to observe diverse organizational strategies across multimodal artifacts.
To better balance, we increased the proportion of code-with-output cells to 50\%, resulting in 34 cell-based artifacts (13 code-only, 17 code-with-output, and 4 narrative) and 17 data-based artifacts.
Within the 17 data-based artifacts, we adjusted the number of data tables and visualizations to align with visualization comparison tasks~\cite{liu2019understanding, weinman2021fork}, resulting in 5 data tables and 12 visualizations.

\textbf{Hardware setup.} 
The study was conducted using a Meta Quest 3 headset (2,064 × 2,208 pixels resolution per eye), running our prototype as an application without the need for a tethered PC or wireless streaming.
The untethered setup allowed participants to move freely within a roughly 5x5m physical area. 
At the beginning of each trial, participants were positioned at the center of the workspace. 
\added{Both cell and data-based artifacts were uniformly fixed at 0.3mx0.4m and were not resizable during the study.
We used a fixed artifact size to control for size-driven placement decisions.
All artifacts were 1 meter in front of the participants and initially clustered in random positions (see~\autoref{fig:process}-(1)) to avoid users' organizational strategies influenced by predefined organizations.}

\begin{figure}[t]
    \centering
    \includegraphics[width=1\columnwidth]{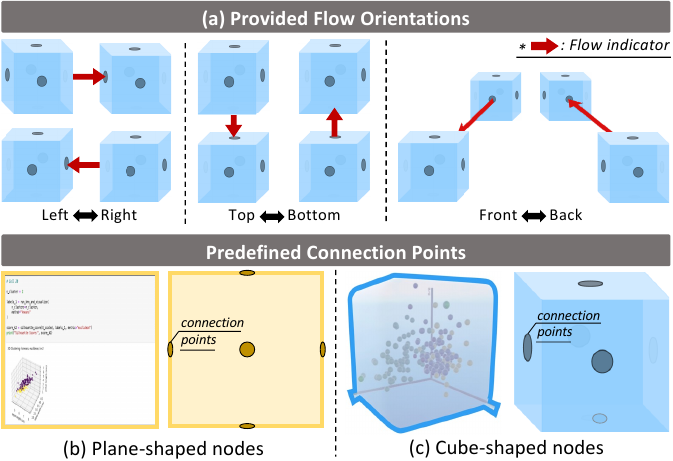}
    \caption{Illustration of the designed mechanism for creating flows. (a) The system provides six orientations, which can be arranged either on (b) a planar surface or within (c) a cubic volume.}  
    \label{fig:flow_mech}
\end{figure}

\begin{figure*}[t]
    \centering
    \includegraphics[width=1\textwidth]
    {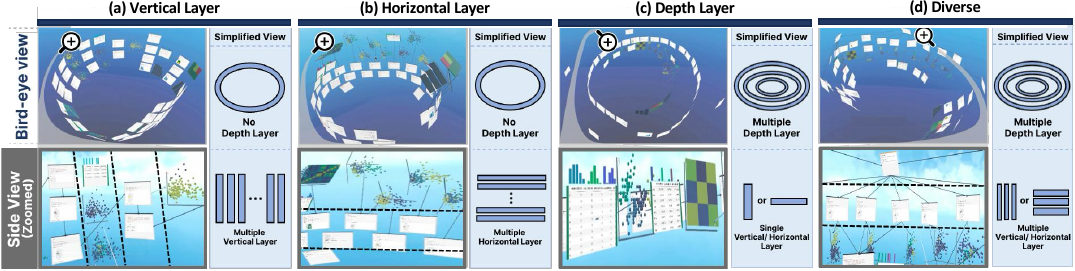}
    \caption{Employed spatial layouts. (a) Artifacts are grouped into multiple vertical layers, while (b) artifacts are grouped into multiple horizontal layers. (c) Artifacts are organized and differentiated by depth, forming multiple ring-like layouts. (d) Participants employed a combination of multiple layouts, using either vertical or horizontal layering while consistently incorporating depth.}    
    \label{fig:placement}
\end{figure*}

\subsection{Study Conditions \& Data}
Based on our study design, we controlled the representation of flow indicators to examine how different flow representations influence organizational strategies. 
Specifically, users often adopt alternative organizational strategies to mitigate occlusion caused by flow representations.
Meanwhile, implicit flow representations are a promising approach for mitigating visual clutter in immersive environments~\cite{elmqvist2007employing, looser2004through}.
For instance, transparency has been demonstrated as an effective technique, as it allows overlapping information to remain partially visible~\cite{diepstraten2002transparency}.
Another potential proxy involves relying solely on spatial cues, without any line-based connectors, since spatial cues in 3D environments can more effectively convey intended reasoning than in 2D settings~\cite{in2023table}. 
\added{While many flow-management techniques exist, such as curved links~\cite{xu2012user} and edge bundling~\cite{sun2015biset, toeda2017convergent}, we aimed to examine how much flow visibility analysts need while avoiding the organization from occlusions. 
Therefore, we focused on transparency as a controlled manipulation of flow representation.}
In summary, the study conditions were defined as follows:

\begin{itemize}[leftmargin=1em]  
    \item \fixedB: Participants were to use only explicit flow indicators (0\% transparency), and all relationships between artifacts had to be clearly represented with explicit flow indicators.
    
    \item \hybridB: Participants could adjust the transparency of flow indicators as needed, choosing among three levels: 0\% (explicit line), 50\% (implicit line), and fully transparent (spatial proxy only). For instance, they could use half transparency to preserve connections while ensuring clear readability, or set an indicator to full transparency and rely only on the spatial proxy.
\end{itemize}

For the data we used for the study, we prepared two separate exploratory data analyses, one using the Iris dataset~\cite{iris} and one using the Wine dataset~\cite{wine}, both involving data loading, basic transformations, and visualization comparisons.
The analysis began with understanding the data structure, supported by data tables and 2D scatter plots as initial overviews. 
We then applied K-Means~\cite{likas2003global} and DBSCAN~\cite{khan2014dbscan} with varying parameter values to demonstrate how different configurations influence the cluster.
The resulting clusters were visualized through 3D scatterplots, allowing participants to inspect relationships and cluster boundaries. 
The analysis concluded with an evaluation comparing K-Means and DBSCAN results using bar charts and heatmaps. 
Additionally, each cell was annotated with its execution order (e.g., Code 1, Code 2).

\begin{table}[t]
    \centering 
    \includegraphics[width=1\columnwidth]
    {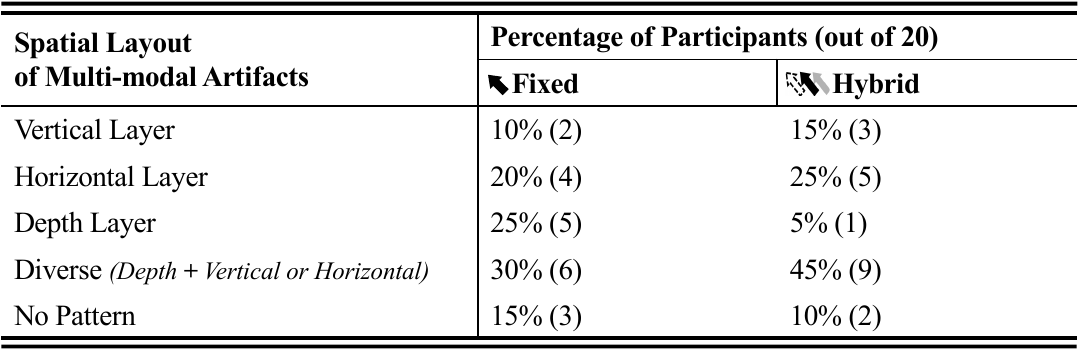}
    \caption{Distribution by spatial layouts in multimodal immersive computational notebooks.}    
    \label{table:table_placement}
\end{table}

\begin{figure*}[t]
    \centering
    \includegraphics[width=1\textwidth]
    {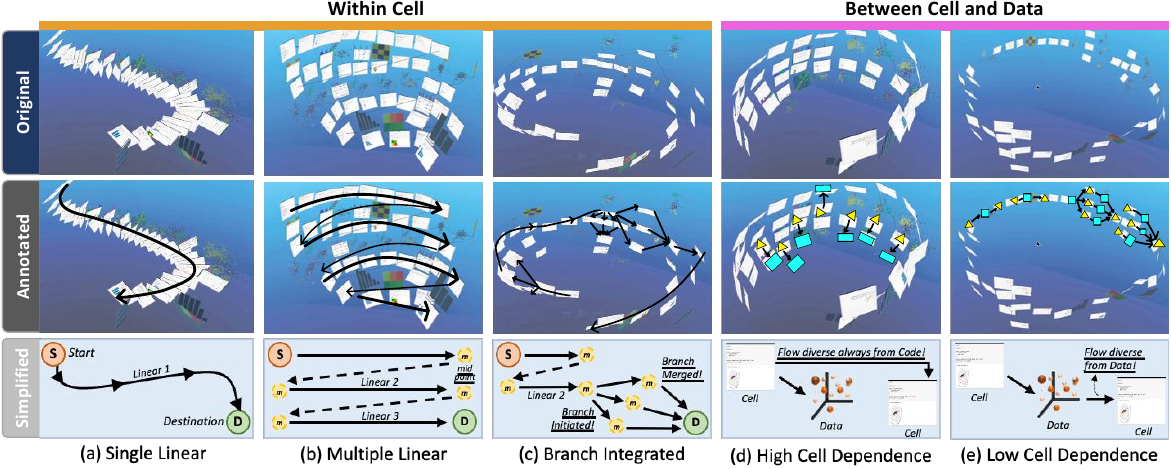}
    \caption{Employed flow. For flows within cells, (a) participants used a single linear flow, in which the flow proceeds in a single direction without changes; (b) flows changed direction multiple times while remaining linear, resulting in multiple linear flows; and (c) participants incorporated branching structures, forming branch-integrated flows. For flows between cells and data, (d) flows were primarily controlled by cells, with no flows originating from data, indicating high cell dependence; whereas in (e) flows also originated from data artifacts, reflecting low cell dependence.}    
    \label{fig:flow}
\end{figure*}

\subsection{Interaction Design}
Participants interacted with artifacts using a hand-held controller. 
A middle trigger was used to grab, move, and place artifacts, while an index trigger was used for selection. 
To create a flow indicator, participants selected two artifacts and chose a direction from a menu attached to the left-hand, which supported six orientations, as shown in~\autoref{fig:flow_mech}-(a).
Indicators were anchored to one of predefined connection points, with plane-shaped nodes for cell-based artifacts (see~\autoref{fig:flow_mech}-(b)) and cube-shaped nodes for data-based artifacts (see~\autoref{fig:flow_mech}-(c)).
To remove the flow indicator, participants selected two artifacts and pressed the thumb button.
Within \hybridT, participants could click an employed flow indicator to adjust its transparency between 0\% (explicit) and 50\% (implicit), while participants could choose not to create an indicator for a fully transparent (spatial proxy only) indicator.

\subsection{Study Procedure \& Task}

The study used a full-factorial within-subjects design, counterbalanced across two groups using a Latin square to mitigate order effects.
Each study session lasted up to an average of 90 minutes, and participants completed the study in the following sequence.
\added{The study protocol was approved by the institutional review board.}

\textbf{Introduction.}
Participants were welcomed, briefed on the purpose of the study, and asked to review and sign the consent form.
Afterward, participants put on the Meta Quest 3 headset and ensured a comfortable fit and clear visuals.

\textbf{Training.}
Participants then completed a training phase.
In training, we used a small set of artifacts, such as five codes, one narrative, two data tables, and two visualizations. 
They practiced grabbing artifacts and placing them in desired positions, and creating/deleting flow indicators. 
We ensured participants were comfortable with the interaction before proceeding to the study task.

\textbf{Task.}
Once participants demonstrated proficiency, they moved on to the study task, which involved organizing a larger number of multimodal artifacts using the same interaction in training. 
\added{To focus the study on organization, the cells were not executable or editable during the task, and each cell-based artifact was annotated with its execution order.
This allowed us to control the analytic content while requiring participants to organize artifacts according to computational relationships.
Participants were instructed to organize the artifacts in a way that made sense to them, while ensuring that the cell-based artifacts followed the correct annotated execution order.
They were also asked to complete the task quickly and accurately; therefore, we can capture organizational efficiency, rather than downstream communication success.
Furthermore, we informed the organization would be shared with other analysts to encourage participants to create an organization that clearly communicated flow, relationships, and analytical intent.}

\textbf{Questionnaires.}
After completing the task, participants filled out the NASA-TLX questionnaires to provide subjective ratings for each condition.
We also collected qualitative feedback to capture their preferences. 
The study concluded after all tasks and questionnaires were completed.

\subsection{Participants}
\added{Following prior immersive computational notebook~\cite{in2024evaluating} and immersive analytics studies~\cite{cordeil2019iatk, davidson2022exploring}, we recruited 20 participants (ages 18--40; 15 male, 5 female) from a university mailing list.}
Eligibility was screened based on prior experience with computational notebooks and data science. 
\added{Participants reported their familiarity with VR: 4 used daily, 4 used it weekly, 8 had experience less than 5 times, and 2 had no prior experience.
All participants had normal or corrected-to-normal vision and volunteered for the study.}

\subsection{Measures}

To understand how participants organized the multimodal artifacts in immersive computational notebooks, we recorded the entire organization process. 
In addition, we collected quantitative measures to address our research questions:
\textbf{Error Scores.}
We recorded an error whenever a participant connected cells in an incorrect execution order.
\textbf{Number of Artifacts Moved.}
We recorded the total number of artifacts that participants moved. 
We further broke this down by artifact type, including codes, narratives, data tables, and visualizations.
\textbf{Completion Time.}
We measured the total time from the start of the task to the moment participants indicated they were finished.
\textbf{Total number of Flow Indicators.}
We computed the total number of flow indicators.
\textbf{Creation and Deletion of Flow Indicators.}
We also logged every instance in which participants created or deleted a flow indicator. 

Using the recordings, we collected the spatial layout across conditions (RQ1) and analyzed the employed flow (RQ2).
Finally, we compared recordings and quantitative measures between \fixedT~and \hybridT~to investigate how the representation of flow indicators influenced participants’ organizational strategies (RQ3).

\section{Results}
\label{sec:results}

This section discusses participants' spatial layouts and employed flows, and reports the quantitative results and subjective ratings.
\added{To categorize the observed spatial layouts and employed flow, two authors independently coded all final organizations using a codebook developed from observed organizational patterns.}
Figures illustrating all final layouts and flows, along with the detailed statistical analysis, are included in the supplementary material.

\subsection{Spatial Layout of Multimodal Artifacts}

From the user study, we identified four distinct spatial layouts: Vertical layer, Horizontal layer, Depth Layer, and Diverse (see~\autoref{fig:placement} and~\autoref{table:table_placement}).
Below, we describe how each spatial layout is justified and report how participants adopted them across conditions.

\textbf{Justification.}
We categorized participants’ layouts based on how they grouped artifacts and how those groups were spatially organized. 
We defined a Vertical Layer when participants form vertically expanded groups and extend their organization by placing additional vertically expanded groups to the left or right (see~\autoref{fig:placement}-(a)).
In contrast, we defined the Horizontal Layer when participants use horizontally expanded groups and extend the organization by placing additional horizontal groups above or below (see~\autoref{fig:placement}-(b)).
We defined a Depth Layer when participants used depth to arrange artifacts along the front–back axis, with only the closest artifacts remaining visible. 
Specifically, participants employed multiple depth layers to differentiate artifact types, such as positioning code, narrative, and data artifacts at distinct depths (see~\autoref{fig:placement}-(c)).
In addition, we defined a Diverse layout when participants combined the Depth Layer with Vertical or Horizontal Layer (see~\autoref{fig:placement}-(d)).
Lastly, we define a No-pattern when participants did not follow a consistent or recognizable grouping strategy.

\textbf{Results.} 
We observed that 2 participants adopted Vertical Layer in the \fixedT~and 3 in the \hybridT, representing a 5\% increase.
Horizontal layering was used by 4 participants in the \fixedT~and 5 in the \hybridT, also reflecting a 5\% increase.
In contrast, Depth layering was used by 5 participants in the \fixedT~but only 1 in the \hybridT, indicating a 20\% decrease.
Diverse layouts were observed in 6 participants in the \fixedT~and 9 in the \hybridT, corresponding to a 15\% increase.
Finally, No-pattern layouts appeared in 3 participants under the \fixedT~and 2 under the \hybridT, showing a 5\% decrease.

\begin{table}[t]
    \centering 
    \includegraphics[width=1\columnwidth]
    {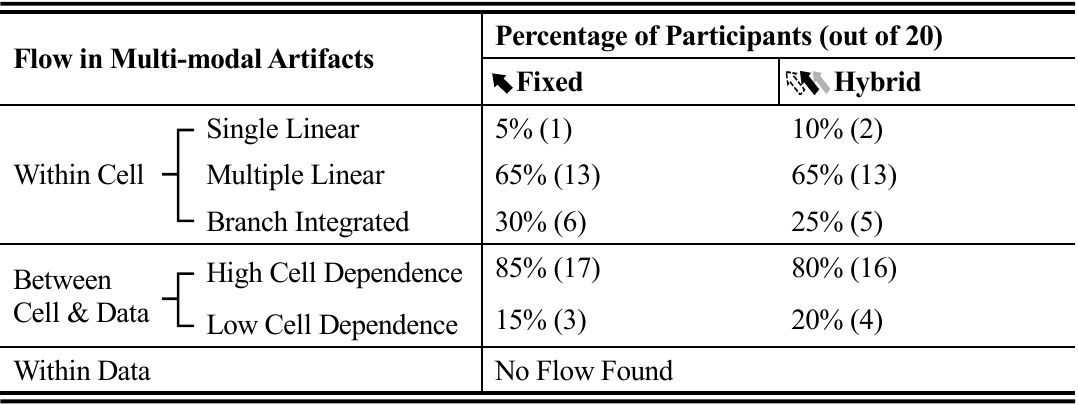}
    \caption{Distribution by data flow in multimodal immersive computational notebooks.}    
    \label{table:table_flow}
\end{table}

\begin{figure}[t]
    \centering
    \includegraphics[width=0.9\columnwidth]
    {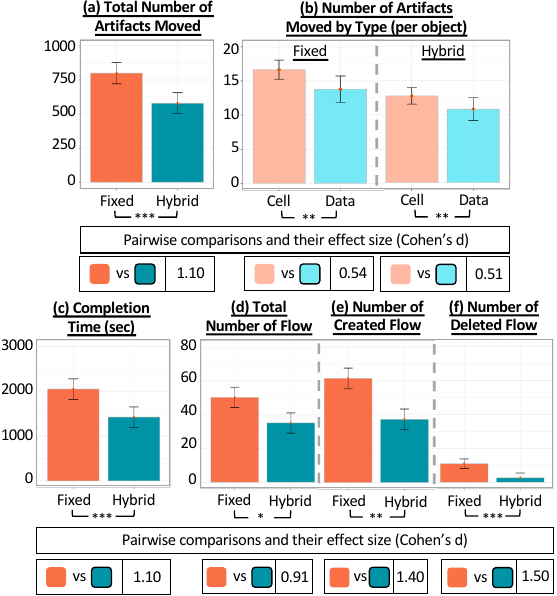}
    \caption{Participant's (a) Total number of artifacts moved to complete the task, and (b) number of artifacts moved by type (cell-based vs. data-based). (c) Total task completion time. (d) Total number of flows employed, computed as (e) the total number of flows created minus (f) the total number of flows deleted. Statistical significance is denoted as *** for p$<$0.001, ** for p$<$0.01, and * for p$<$0.05.}    
    \label{fig:quant}
\end{figure}

\begin{figure}[t]
    \centering
    \includegraphics[width=1\columnwidth]
    {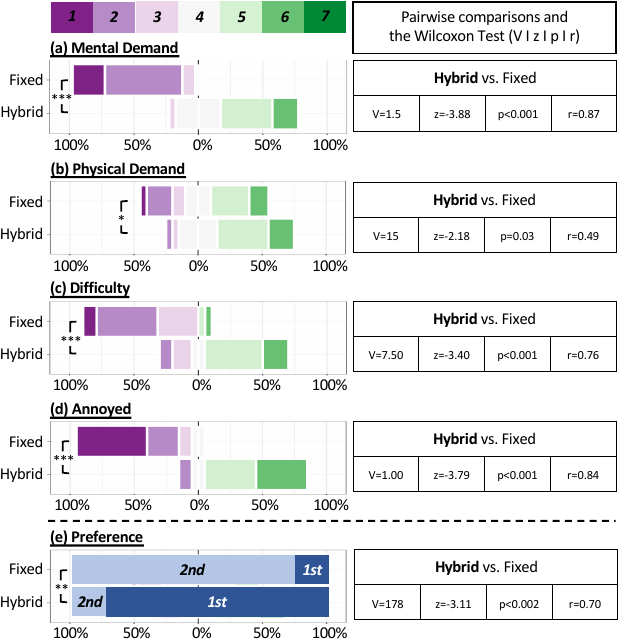}
    \caption{Participant subjective ratings. (a) Mental demand, (b) physical demand, (c) perceived difficulty, (d) annoyance, and (e) preference. Statistical significance is denoted as *** for p$<$0.001, ** for p$<$0.01, and * for p$<$0.05.}    
    \label{fig:rating}
\end{figure}

\subsection{Flow in Multimodal Artifacts}

From the user study, we identified five distinct flows: Single Linear, Multiple Linear, Branch-Integrated between cell-based artifacts, and High and Low Cell Dependence between cell and data-based artifacts (see~\autoref{fig:flow} and~\autoref{table:table_flow}).
Below, we describe how each flow is justified and report how participants adopted them.

\textbf{Justification.}
We categorized flows based on how participants expressed relationships among cell-based artifacts and data-based artifacts.
\added{For participants who employed fully transparent flow indicators, we asked them in the post-study interview to describe their intended flow among artifacts.}
For flows between cell-based artifacts, we considered flow as an execution order, following the execution-order schemas identified in a user study on a 2D computational notebook~\cite{harden2022exploring}. 
We defined Single Linear as an execution order that proceeds in one direction (e.g., only left$\rightarrow$right) from start to end without changing direction (see~\autoref{fig:flow}-(a)).
In contrast, we defined Multiple Linear when participants introduce additional linear execution by changing direction, such as starting a new row or column (see~\autoref{fig:flow}-(b)). 
We further defined Branch-Integrated, in which participants introduced branching or tree-like execution orders, indicating divergent execution paths for effective comparisons (see~\autoref{fig:flow}-(c)).
For flows between cells and data-based artifacts, categorization was driven by how participants perceived the dependency of cell execution. 
We defined High Cell Dependence, in which flow indicators are never initiated from data artifacts.
In these scenarios, the execution flow should return to a cell after visiting data (see~\autoref{fig:flow}-(d)). 
In contrast, Low Cell Dependence was observed when participants also initiated flows from data artifacts toward cells, allowing flow to proceed through data to subsequent cells (see~\autoref{fig:flow}-(e)).
Finally, we note that no flows were observed between data artifacts.

\textbf{Results.}
We observed that between cell-based artifacts, Single Linear was adopted by 1 participant in the \fixedT~and 2 in the \hybridT, representing a 5\% increase.
Multiple Linear were adopted by 13 participants in both the \fixedT~and the \hybridT.
Branch-Integrated were adopted by 6 participants in the \fixedT~and 5 in the \hybridT, corresponding to a 5\% decrease.
Between cell and data-based artifacts, High Cell Dependence was observed in 17 participants in the \fixedT~and 16 in the \hybridT, representing a 5\% decrease. 
In contrast, Low Cell Dependence was adopted by 3 participants in the \fixedT~and 4 in the \hybridT, showing a 5\% increase.
Finally, we observed that no participants employed flows between data artifacts in either condition.

\subsection{Quantitative Results}

In this section, we report results from quantitative measures that highlight performance differences between conditions (see~\autoref{fig:quant}).

\textbf{Error Score.}
The provided annotations in cell-based artifacts enabled participants to complete the tasks accurately across all conditions, resulting in no variance in accuracy.

\textbf{Number of Artifacts Moved.}
We observed a significant effect of condition on the total number of artifacts moved (***). 
The linear mixed-effects model showed a significant main effect of condition, \textit{$\beta = 0.36$, $\mathrm{SE} = 0.09$, $\chi^2(1) = 12.17$, $p < .001$}.
Pairwise comparisons indicated that participants moved significantly fewer artifacts in the \hybridT~(avg. 579.8, CI = 98.56) than in the \fixedT~(avg. 800.2, CI = 111.15), \added{with a large effect size (\textit{Cohen's $d = 1.10$})}.
We further examined the number of moved artifacts by artifact type within each condition. 
In the \fixedT, participants moved significantly more cell-based artifacts than data-based artifacts (**), \textit{$\chi^2(1) = 52.34$, $p < .001$} (cell-based avg. = 16.64, CI = 2.25; data-based avg. = 13.79, CI = 2.68), \added{with a medium effect size (\textit{Cohen's $d = 0.54$})}. 
Similarly, in the \hybridT, participants also moved significantly more cell-based artifacts than data-based artifacts (**), \textit{$\chi^2(1) = 8.29$, $p = .004$} (cell-based avg. = 12.76, CI = 1.44; data-based avg. = 10.81, CI = 2.05), \added{with a medium effect size (\textit{Cohen's $d = 0.51$})}.

\textbf{Completion Time.}
Completion time was significantly affected by the condition (***).
A linear mixed-effects model revealed a significant main effect of condition, \textit{$\beta = 1.80$, $\mathrm{SE} = 0.16$, $\chi^2(1) = 52.34$, $p < .001$}.
Pairwise comparisons showed that participants completed significantly faster in the \hybridT~(avg. 1423.56s, CI = 225.15s) than in the \fixedT~(avg. 2047.65s, CI = 301.50s), \added{with a large effect size (\textit{Cohen's $d = 1.10$})}.

\textbf{Total number of Flow Indicators.}
The total number of flow indicators was marginally influenced by the condition (*). 
The linear mixed-effects model showed \textit{$\beta = 1.03$, $\mathrm{SE} = 0.38$, $\chi^2(1) = 6.63$, $p = .010$}. 
Participants had a higher number of flow indicators in the \fixedT~(avg. 50.35, CI = 6.40) compared to the \hybridT~(avg. 35.15, CI = 8.94), \added{with a large effect size (\textit{Cohen's $d = 0.91$})}.

\begin{figure*}
    \centering
    \includegraphics[width=1\textwidth]
    {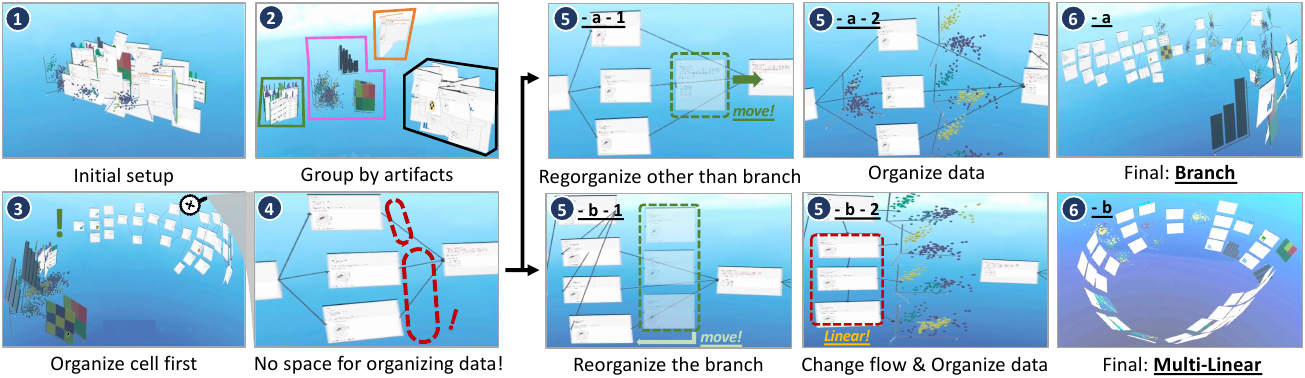}
    \caption{Illustration of the organization process from the (1) initial setup to the (6-a / 6-b) final arrangement. The figure highlights two strategies used to ensure free space for organizing data: participants who (5-a-1) reorganized branches, or (5-b-1) did not reorganize branches.} 
    \label{fig:process}
\end{figure*}

\textbf{Creation and Deletion of Flow Indicators.}
The number of flow indicator creation (**) and deletion (***) was significantly influenced by the condition.
For creation of flow indicators, the linear mixed-effects model showed \textit{$\beta = 1.21$, $\mathrm{SE} = 0.39$, $\chi^2(1) = 8.48$, $p = .0036$}. 
Participants created more in the \fixedT~(avg. 61.35, CI = 6.78) compared to the \hybridT~(avg. 37.20, CI = 9.17), \added{with a large effect size (\textit{Cohen's $d = 1.40$})}.
For deletion of flow indicators, the linear mixed-effects model showed \textit{$\beta = 1.58$, $\mathrm{SE} = 0.34$, $\chi^2(1) = 17.73$, $p < .001$}. 
Participants deleted more in the \fixedT~(avg. 11.00, CI = 3.54) compared to the \hybridT~(avg. 2.65, CI = 1.06), \added{with a large effect size (\textit{Cohen's $d = 1.50$})}.

\textbf{Subjective Ratings and Preference.}
\added{Although we used the original NASA-TLX questionnaires, we report only the measures that showed significant results. The comprehensive results are provided in the supplementary material.}
We found a significant effect of condition on mental demand (***), physical demand (*), difficulty (***), annoyed (***), and preference (**) (see~\autoref{fig:rating}).
Participants rated the \fixedT~as more mentally demanding (avg. 1.95, CI = 0.36) than the \hybridT~(avg. 4.75, CI = 0.40), \added{with a large effect size (\textit{$r = 0.87$})}.
Similarly, participants rated the \fixedT~as more physically demanding (avg. 3.95, CI = 0.72) than the \hybridT~(avg. 4.65, CI = 0.49), \added{with a medium effect size (\textit{$r = 0.49$})}.
In addition, participants considered the \hybridT~(avg. 3.85, CI = 0.99) easier to use than the \fixedT~(avg. 2.60, CI = 0.86), \added{with a large effect size (\textit{$r = 0.76$})}.
\added{Furthermore, participants felt that the \fixedT~was more annoying to use (avg. 1.75, CI = 0.48) than the \hybridT~(avg. 5.00, CI = 0.57), also with a large effect size (\textit{$r = 0.84$}).}
\added{Finally, the \hybridT was rated as more preferable (17 out of 20), with a large effect size (\textit{$r = 0.70$}).}

\section{Key Findings \& Discussion}
\label{sec:discussion}

\textbf{Artifacts tended to be organized into two dominant groups: cell and data-based artifacts.}
While 4 participants in the \fixedT~and 3 in the \hybridT~organized artifacts into more fine-grained groups---code, narrative, and data---the majority of participants (13 in the \fixedT~and 15 in the \hybridT) grouped artifacts into two higher-level categories: cell and data-based artifacts.
In this grouping, code and narratives were treated as cell-based artifacts, while data tables and visualizations were grouped as data-based artifacts. 
We believe this aligns with the law of proximity in Gestalt theory~\cite{wertheimer1938gestalt}, where participants tended to group functionally and conceptually related artifacts.
This is likely because code and narratives are encapsulated within cell-based structure and jointly express procedural logic and explanatory context~\cite{rule2018exploration}, whereas data tables and visualizations represent instantiated data and primarily support result inspection and interpretation.
Therefore, code and narrative artifacts were frequently organized together, and similarly, data tables and visualizations were often located closely.

\textbf{Half of the participants adopted a depth layer to organize a multimodal immersive computational notebook.}
Similar to prior organizational studies in immersive environments~\cite{in2025exploring, luo2022should}, we observed that the majority of participants organized in a circular layout (16 for both \fixedT~and \hybridT).
We further observed that participants largely favored using the depth to organize multimodal artifacts (11 participants in the \fixedT~and 10 in the \hybridT), whereas participants with single-modal artifacts generally organized without the use of depth~\cite{davidson2022exploring, in2023table, lisle_sensemaking_2021}.
Meanwhile, depth-based organization has been shown to serve multiple purposes, including conveying hierarchical relationships and lowering the perceived priority of artifacts~\cite{elmqvist2009hierarchical, robertson1991cone}.
We observed similar behaviors in our study, with 5 participants for both conditions placing narrative cells in the outer layer, indicating that narratives were considered less important.
In addition, participants who prioritized visibility positioned narrative cells toward the top or bottom of the layout; however, narrative cells were still consistently arranged along the outer layer.
One participant explicitly mentioned this, stating, ``I put the narrative cells behind as they would be used less. (P7)''

\textbf{Cell serves as a spatial anchor for organizing multimodal artifacts.}
In our study, participants were asked to reorganize artifacts from an initial cluster (see~\autoref{fig:process}-(1)). 
During this process, we observed that 60\% of participants (12 out of 20) began by organizing cell-based artifacts, carefully adjusting spacing and alignment (see~\autoref{fig:process}-(3)). 
Although these careful adjustments resulted in more frequent interactions with cell-based artifacts (see~\autoref{fig:quant}-(b)), participants did not move the cells or made only minor adjustments when they proceeded to organize data-based artifacts.
This behavior was observed among a large proportion of participants who organized cells first (8 out of 12).
By fixing the spatial configuration of cells first, participants were able to integrate data artifacts while maintaining a coherent representation of analytical structure and execution order.
We believe this behavior is evidence that cell-based artifacts serve as stable reference points, helping preserve spatial positioning and relationships across multimodal artifacts~\cite{hubenschmid2022relive, muller2017remote}. 

\added{\textbf{Flows predominantly initiate from cell-based artifacts.}}
We identified that nearly all participants (17 out of 20) consistently initiated flows only from cell-based artifacts. 
From in-depth analysis, we found that this behavior potentially reflects how participants interpret the relationship in multimodal immersive computational notebooks. 
Specifically, we considered that participants perceived the data artifacts as outcomes or side branches derived from code execution. 
As a result, participants tended to express computational sequencing primarily through cell-to-cell flows, even when it reduced visual clarity.
For instance, in P10’s organization, even when data artifacts were placed between two cells, the flow did not initiate from the data artifact; instead, it initiated from the preceding cell (see~\autoref{fig:flow}-(d)). 
Consequently, the flow indicator passed across the data artifact, introducing visual clutter and requiring participants to revisit the cell before proceeding to the next cell. 
In contrast, when flows were also initiated from data-based artifacts, all participants (3) produced more continuous flow paths that better integrated code and data (see~\autoref{fig:flow}-(e)). 
This observation aligns with prior findings suggesting that users are often willing to tolerate a certain level of discomfort to achieve their intended goals~\cite{andrews2010space}.
P10 further supports this observation by commenting, ``I think data and code should not be treated as a continuous sequence.''

\begin{figure}[t]
    \centering
    \includegraphics[width=0.85\columnwidth]
    {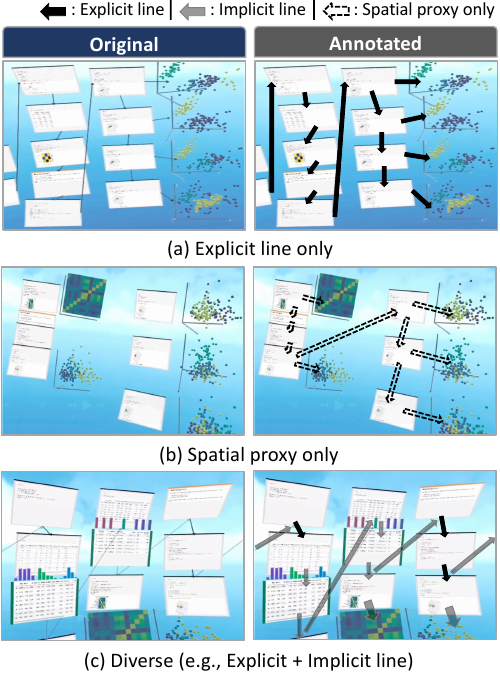}
    \caption{Illustration of how flows were utilized in the \hybridT. (a) Participants used only explicit lines to represent flows, while in (b) participants chose not to use any lines to indicate flows. In contrast, (c) participants employed multiple strategies, combining explicit and implicit lines. No participants used implicit lines only.}    
    \label{fig:flow_hybrid}
\end{figure}

\added{\textbf{Execution order tended to become linear when multimodal artifacts were present.}}
Prior work on extending coding interfaces into spatial environments, such as 2D desktop~\cite{harden2023sage3} and 3D VR~\cite{in2024evaluating}, has shown that participants dominantly organize code in non-linear order, as non-linearity more intuitively reflects the nature of complex data analysis workflows~\cite{deline2012debugger}. 
\added{Although our study also involved non-linear execution orders (e.g., branch), we observed that relatively smaller participants adopted non-linear execution orders (30\% in \fixedT~and 25\% in \hybridT).
We identified participants who began by organizing cell-based artifacts as a key factor contributing to this shift. 
To be specific, four participants first established a non-linear execution order in both \fixedT~and \hybridT. 
However, as they proceeded to organize data artifacts, they appeared to prioritize creating sufficient free space around the corresponding cells to avoid overlap.
Although branching structures could have been maintained by repositioning other cells (see~\autoref{fig:process}-(5-a-1)), these participants reorganized their branches into multiple linear execution orders instead (see~\autoref{fig:process}-(5-b-1)).
One participant explained this change by stating, ``I have changed my organization (branch to linear) because of the occlusion issues. (P12)''}

\textbf{\fixedB~incurs higher interaction effort due to frequent flow adjustments to maintain visual clarity.}
We observed that participants in the \fixedT~interacted with flow indicators more frequently, with 39\% more creation and 76\% more deletion compared to the \hybridT. 
We considered that this is due to the additional effort required to manage visual clutter in the \fixedT.
\added{Specifically, we observed that when flow indicators introduced visual clutter, they often first attempted to reposition artifacts to identify a place that could reduce clutter.}
However, such placements were difficult to find, requiring participants to delete and create flow indicators with a different orientation. 
\added{We believe this iterative process contributed to the 38\% higher number of artifact movements in the \fixedT.}
Furthermore, we believe additional interaction burden contributed to the 44\% slower task completion in the \fixedT.
\added{However, in the \hybridT, the issue was mitigated as it allows for adjusting the flow representation, such as switching to implicit lines.}
One participant explicitly noted, ``I used the implicit line for the ones that go across other artifacts to avoid visual clutter. (P16)''
This reduced interaction overhead is also reflected in participants’ subjective ratings: they reported the \fixedT~as more difficult to finish the task and annoying to organize, and consequently expressed a stronger preference for the \hybridT.

\textbf{\hybridB~enables participants to establish a flow to be more informative with less visual clutter.}
\added{Although we did not observe distinct differences in spatial layout between the \fixedT~and \hybridT, all participants in the \hybridT~employed multiple flow representations.
Specifically, 70\% of participants used explicit lines for flows between cell-based artifacts, while 55\% used implicit lines for flows between cell and data-based artifacts.
Implicit lines were often used to reduce visual clutter, especially when flows crossed other artifacts (see~\autoref{fig:flow_hybrid}-(c)).
In contrast, 30\% of participants did not create any flow indicators; instead place related artifacts in close spatial proximity to convey semantic relationships (see~\autoref{fig:flow_hybrid}-(b)).
The absence of flow indicators between cells and data-based artifacts was also observed when participants assigned low priority to data artifacts.}
This aligns with prior studies highlighting the role of flow representations in establishing and communicating flows~\cite{figl2024guiding}.
One participant explicitly emphasized this semantic differentiation; ``I used implicit lines to differentiate code and data, and I did not use lines for narrative cells. (P15)''

\begin{table}[t]
    \centering 
    \includegraphics[width=0.95\columnwidth]
    {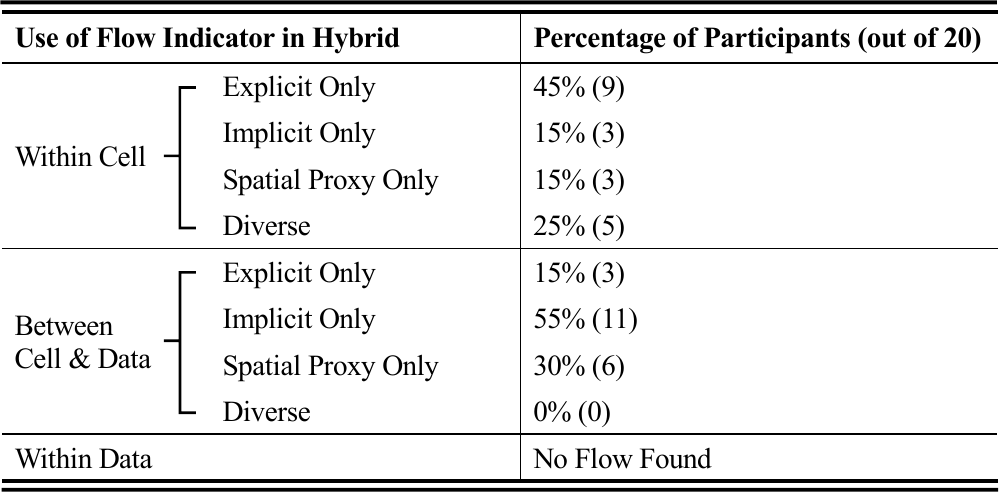}
    \caption{Distribution by the use of flow indicators in \hybridT.} 
    \label{table:table_flow_hybrid}
\end{table}

\section{Design Implications}
\label{sec:implication}

\added{\textbf{Design Implication 1: Consider depth as a potential organizational dimension and support depth-aware interaction.}
Unlike prior organizational studies on single-modal artifacts, which primarily reported single-layer layouts~\cite{in2025exploring}, we observed that over half of the participants leveraged depth to organize multimodal artifacts to separate cell and data-based artifacts.
Furthermore, participants used depth to encode semantic distinctions, such as placing narrative cells in outer layers to indicate lower priority.
We considered that these observations suggest that depth can serve as a potential organizational dimension where complex relationships exist.
However, depth-based organization may also introduce additional effort~\cite{in2025towards}, such as requiring users to take extra steps to view artifacts occluded by closer artifacts or to move closer to inspect artifacts placed in outer layers.
Therefore, systems may benefit from depth-aware interaction mechanisms (e.g., piling~\cite{bach2015small}) to support navigation and interaction with depth-oriented layouts.}

\added{\textbf{Design Implication 2: Support cell-centered organization for multimodal immersive computational notebooks.}
Data analysis is inherently dynamic~\cite{rule2018exploration}, meaning that organizations may continuously evolve as analysts add new cells or data artifacts.
Meanwhile, our observations suggest that cells may serve as spatial anchors that shape the overall structure of organizations.
Specifically, most participants began by arranging cells and then positioned data artifacts around the corresponding cells, making only minor adjustments to the cell layout.
Therefore, we suggest cell-centered organization mechanisms that help preserve the stability of previously organized cells while allowing data artifacts to be reorganized in response to changes in cell organization.
Such support may help maintain workflow stability and reduce cognitive disruption caused by frequent structural changes.}

\added{\textbf{Design Implication 3: Support selective and context-aware flow representations.}}
Prior studies have shown that providing various flow representations allows analysts to choose encodings that best match their analytical intent~\cite{gansner2006improved, wong2003edgelens}. 
\added{Similarly, our observations suggest that multiple levels of transparency can allow analysts to embed different purposes into their flow.
Specifically, explicit lines were primarily used to express strong relationships, whereas implicit lines were often used to indicate secondary dependencies while mitigating visual clutter.
In addition, participants omitted flow indicators and relied on spatial proximity to communicate semantic relationships.
These behaviors suggest that participants did not treat flows only as a mechanism for connecting artifacts.
Instead, they used flows to convey semantic information, such as task context, artifact priority, and the need to manage visual clutter.
Therefore, systems may benefit from selective, context-aware flow representations that allow users to adjust flow visibility and encoding, rather than mandatory explicit representation.}

\section{Limitation \& Future Work}
\label{sec:limitation}

\added{We identified several limitations that should be examined in broader contexts to assess the generalizability of our findings.
First, participants often grouped code and narrative artifacts, which may reflect the computational notebooks' structure where code and narratives are encapsulated within cells~\cite{knuth1984literate}.
During the pilot study, we did not observe a consistent pattern in which all cell-based artifacts were grouped together, making it difficult to identify this issue earlier and include targeted probes in the interview protocol.
Furthermore, flows predominantly originated from cell-based artifacts rather than data-based artifacts, likely because data outputs were perceived as supporting cell execution.
Therefore, further exploration is needed in more generalized contexts where code, narrative, and data artifacts are not constrained by notebook cell structures.}

\added{In addition, our study used fixed artifact sizes, whereas immersive analytics often involves various artifact sizes~\cite{luo2022should}.
As variable artifact sizes may influence organizational strategies, future work should examine how resizable artifacts affect organizational strategies in multimodal immersive computational notebooks.
Also, participants' prior experience with spatially organizing complex information may have influenced organizational strategies.
In addition, the organization task was designed for early-stage organizational decision-making. 
Future work therefore should examine organization in longer-term, open-ended analytical workflows with larger and more expert participants to improve the generalizability of our findings.}

\added{Finally, although participants used all three transparency levels to externalize semantic meaning or reduce visual clutter, transparency is only one dimension of visual encoding.
Other visual proxies, such as line width, color, and dash patterns, could also convey different relationships or semantic distinctions~\cite{cui2008geometry, dickerson2002confluent}.
Future work should explore how these visual proxies support relationships in multimodal immersive workflows.}

\section{Conclusion}
\label{sec:conclusion}

We explored how analysts organize multimodal artifacts---code, narratives, data tables, and visualizations---in immersive computational notebooks, and how flow representations influence these organizational strategies. 
We found that participants consistently grouped artifacts into cell-based and data-based and frequently leveraged the depth dimension to separate these groups.
In addition, flows almost always originated from cell-based artifacts, indicating that cell-based artifacts were treated as the primary proxy during organization. 
Furthermore, participants used cell-based artifacts as spatial anchors when organizing their workspaces. 
However, we believe that our findings, where spatial layout and flows were predominantly governed by cell-based artifacts, were potentially influenced by the nature of computational notebooks. 
Therefore, further exploration with more distinct representations is needed to assess the broader generalizability of our findings.

\acknowledgments{This work was partially supported by the Laboratory for Analytic Sciences at North Carolina State University through an NSA/DoD subcontract and the National Science Foundation under Grants IIS-2544317, TI-2612077, and IIS-2441310, and this work was supported by a grant from Kyung Hee University in 2026 (KHU-20262258).}


\bibliographystyle{abbrv-doi}
\bibliography{main}

\end{document}